# Security Enhancement of Quantum Noise Stream Cipher Based on Probabilistic Constellation Shaping

Sheng Liu [(1)]†, Shuang Wei [(2)]†, Wei Wang [(2)]†, Chao Lei [(2)], Tianhe Liu [(2)], Yajie Li [(2)], Yunbo Li [(1)], Dawei Ge [(1)], Dong Wang [(1)], Yongli Zhao [(2)], Dechao Zhang [(1)], Han Li [(1, *)], and Jie Zhang [(2, *)]

[(1)] Department of Fundamental Network Technology, China Mobile Research Institute, Beijing 100053, China, lihan@chinamobile.com
[(2)] School of Electronics Engineering, Beijing University of Posts and Telecommunications, Beijing 100876, China, jie.zhang@bupt.edu.cn

**Abstract** *We propose a QNSC pre-coding scheme based on probabilistic shaping of the basis, to reduce the probability of ciphertext bits that are easier to be intercepted. Experiment results show this scheme can improve the security performance by 100% in terms of Eve's cipher text BER. ©2023 The Author(s)*

## Introduction

With the development of cyber-attack methods, the security issues of optical communication networks have gained more and more attention [1]. Quantum noise stream cipher (QNSC) is a popular physical-layer security approach, and it maps the original signal symbols into a very large symbol constellation with the protection of the quantum noise of optical signals [2].

QNSC was intensively studied from the perspectives of transmission and security performance [3, 4]. In terms of security, there are generally two ways to enhance the security, strengthening the noise effect of the system and reducing the adjacent signal Euclidean distance [5]. In conventional QAM/QNSC systems, the ciphertext symbols follow a uniform distribution in the constellation diagram, i.e., symbols are distributed with the same probability. However, QNSC does not provide absolute security. As we know, according to the encryption process of QNSC, the plaintext is encrypted into the low-order ciphertext with XOR operation, and the low-order ciphertext is further appended on top of the basis, forming a high-order ciphertext symbol. After being masked by quantum noise, the bit error ratio (BER) of the lower bits (the basis) of the high-order ciphertext is relatively high, while the BER of the higher bits (the original low-order ciphertext) of the high-order ciphertext is relatively low. In conventional QAM/QNSC, the low BER for Eve means higher probability of being successfully intercepted. Therefore, the distribution of the lower bits (the basis) is of great importance for security performance. Meanwhile, the probabilistic constellation shaping (PCS) technique is emerging, and it is capable of customizing the probabilistic distribution of symbols in the constellation diagram [6, 7]. Therefore, it could be interesting to study how to improve the security of QAM/QNSC, facilitated by PCS technology, without breaking the original QNSC.

In this paper, we propose a PCS-based pre-coding scheme to modify the distribution of the basis and plaintext, to reduce the BER of low-order ciphertext. We implemented the proposed scheme in a QAM/QNSC testbed. From the perspective of security, experimental results show that this scheme improves the system security performance by about 100% in terms of Eve's BER of low-order ciphertext. From the perspective of transmission performance, experimental results show that this scheme improves OSNR 0.5 dB and 0.6 dB compared with traditional QNSC in back-to-back (BTB) and 160 km fiber.

## The scheme of PCS-based Pre-coding

In an original QNSC system, the plaintext bit is encrypted by the lowest bit of the basis with XOR operation, and the XOR result is further placed at the highest bit over the basis, forming a high-order ciphertext symbol. If we use PCS to shape the ciphertext symbols directly, the order of bits of the symbol is changed [8]. Here, we proposed a QNSC pre-coding scheme with a two-step PCS to customize the distribution of ciphertext symbols and enhance security, as illustrated in Fig. 1.

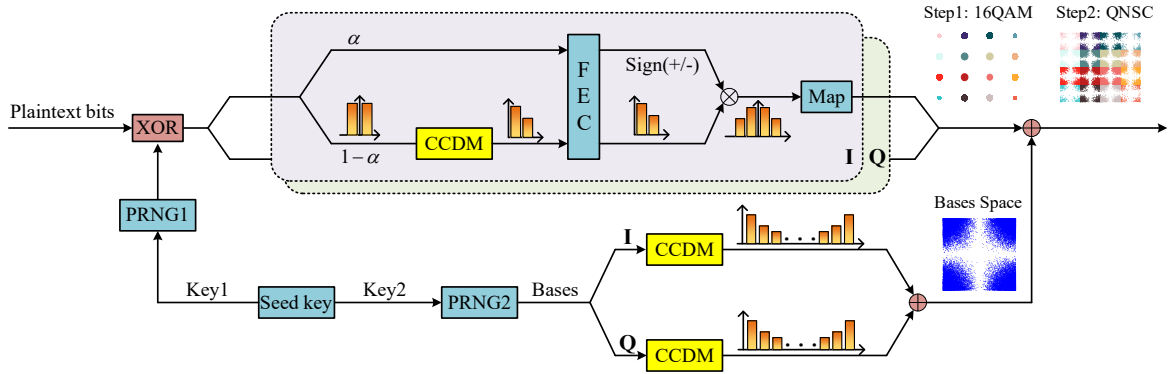

**Fig. 1:** PCS-based plaintext and basis pre-processing scheme for QAM/QNSC

- Pre-coding for basis

In the first step, the seed key is divided into two parts (i.e., key1 and key2) for XOR operations and basis pre-coding. Based on key-2, bases are generated by the pseudorandom number generator (PRNG) and injected into the CCDM module. The basis's probability distribution is changed from uniform to non-uniform. In this paper, the inverse model of Gaussian distribution is adopted.

- Pre-coding for both basis and plaintext

Based on the shaped basis, we can further customize the distribution of the plaintext. After key1 is expanded by PRNG1, each plaintext bit is encrypted with key1 by XOR operation to obtain the ciphertext bit. Following the PCS principle, the ciphertext bits on the transmitter side are split into two branches, with the corresponding proportions of data of $\alpha$ and $1-\alpha$. The data in the lower branch is shaped by a constant composition distribution matching (CCDM), and the output from CCDM follows a desired distribution. The output of CCDM and the data in the upper branch are sent to the FEC encoding modules for generating parity bits. The upper branch bits from the FEC are used as the $\pm 1$ signs, which are multiplied with the shaped data in the lower branch as the amplitude. After the two symbols are superimposed, low-order ciphertext symbols (e.g., 16QAM) with desired distribution are generated.

After getting the shaped basis and ciphertext, the low-order ciphertext bits are appended on top of the basis, forming the high-order ciphertext symbols according to Y-00 protocol. The essence of this step is to attach a random perturbation to each low-order ciphertext symbol within the decision region of itself. The ciphertext symbols near the decision threshold of the low-order ciphertext symbol are easy to cross the decision threshold under the influence of noise, which will cause BER for low-order ciphertext of Eve without the key. The benefit of the scheme is that the ciphertext symbol distribution is specifically designed to increase the BER of the low-order ciphertext of Eve, by increasing the symbol points near the decision threshold of the low-order ciphertext symbol and thus improve the security of QNSC systems.

**Experimental Setup and DSP Flows**

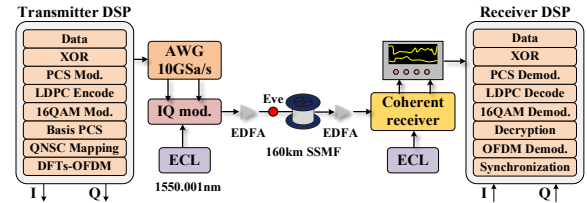

**Fig. 2:** Experimental setup.

To evaluate the performance of the proposed algorithm, we conduct an experiment with our lab-grown setup, as shown in Fig. 2. The output probabilistic distribution from the PCS module is set as $P(1) = 0.68$ and $P(3) = 0.32$ ($H = 3.8$). In the setup above, the plaintext bits are encrypted by key1 and fed into the PCS module, and they are mapped to 16QAM symbols with desired distribution. The 16QAM symbols are further extended by the basis to the high-order ciphered symbols following the Y-00 protocol. OFDM signal is generated and fed into an arbitrary waveform generator (AWG) with 10GSa/s. An external cavity laser (ECL) is used to generate an optical carrier at the wavelength of 1550 nm and power of 6 dBm. After being amplified by the transmitter side erbium doped fiber amplifier (EDFA), the optical power injected into the standard single-mode fiber (SSMF) is 0dBm. The fiber length is 160km. At the receiver, the optical signal is coherently detected. The receiver decrypts the QAM symbols with the pre-shared seed key and further recovers the data by performing QAM demodulation, PCS demodulation, and XOR.

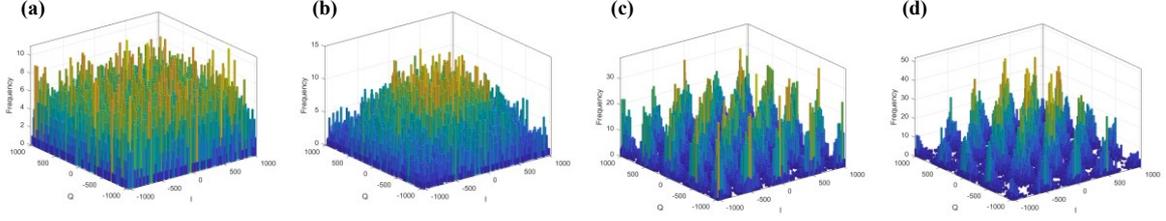

**Fig. 3:** The distribution of constellation points of (a) traditional QNSC, (b) QNSC based shaped 16QAM (c) QNSC based on shaped basis, (d) QNSC based on shaped 16QAM and basis.

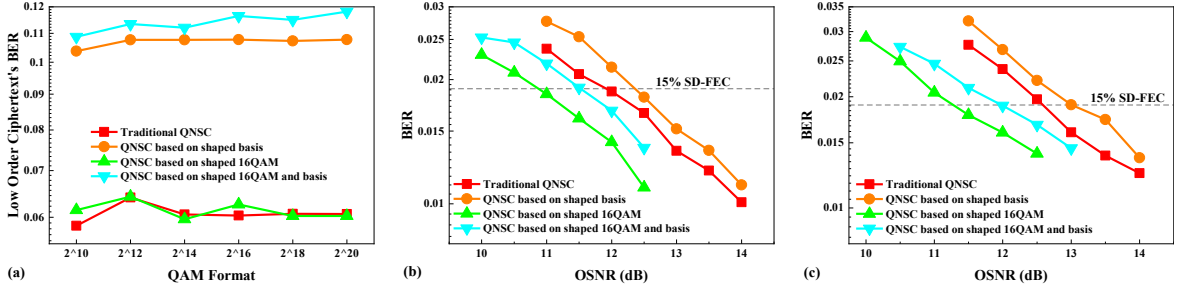

**Fig. 4:** Experimental results (a) low-order ciphertext's BER for Eve; (b) the pre-FEC BER of BTB and (c) 160km fiber

## Experimental Results

In the above-mentioned setup with back-to-back or SSMF, we measured the security and transmission performance in terms of Eve's low-order ciphertext BER and the legal party's pre-FEC BER. To demonstrate the effect of our scheme, three benchmarks are measured: traditional QNSC, QNSC based on shaped 16QAM and QNSC based on shaped basis. The distribution of the constellation points for our scheme and the three benchmarks are shown in Fig. 3. The I and Q axes represent the amplitude of QNSC symbols. Fig. 4 (a) compares Eve's low-order ciphertext BER under different QAM formats. Note that, Eve has the best condition for intercepting information when the signal received by Eve is minimally affected by noise. Hence, the position right after the EDFA in the transmitter is the optimal eavesdropping point for Eve, as shown in Fig. 2. By adjusting the symbol rate, we maintain the net bit rate of the four schemes. It is clearly concluded that our scheme has the highest Eve's low-order ciphertext BER of more than 11%, while traditional QNSC and QNSC based on shaped 16QAM schemes are only about 6%. Our scheme improves security by 100%. The reason is that more symbols cross the decision threshold of 16QAM under the influence of noise in our scheme. It indicates that the system is more resistant to Eve's interception. In the practical splitting attack scenario, Eve will detect a much higher low-order ciphertext BER.

In Fig. 4(b) and (c), we record the pre-FEC BER of four different schemes in two scenarios - BTB and 160 km fiber. Among them, QNSC based on the shaped 16QAM scheme has the best BER performance. The QNSC based on the shaped basis scheme has the highest BER because the minimum Euclidean distance decreases under the same power. In fact, our scheme improves OSNR 0.5 dB and 0.6 dB compared with traditional QNSC in BTB and 160 km fiber under 15% overhead SD-FEC.

## Conclusions

We propose a pre-coding QNSC scheme based on the probabilistic constellation shaping for both of the plaintext and the basis to make the distribution of ciphertext symbols more resistant to eavesdroppers. Experimental results verified that this scheme can improve the security in terms of Eve's low-order ciphertext BER and improve the transmission performance in terms of OSNR by 0.5 dB and 0.6 dB over BTB and 160 km fiber, respectively.

**Acknowledgements**

This work is supported in part by the National Natural Science Foundation of China (NSFC, under grants: 62101063, 61831003, 61901053), and BUPT-CMCC Joint Innovation Center. Sheng Liu, Shuang Wei and Wei Wang contributed equally to this work.

**References**

[1] X. Gong, Q. Zhang, X. Zhang, R. Xuan and L. Guo, "Security Issues and Possible Solutions of Future-Oriented Optical Access Networks for 5G and Beyond," in *IEEE Communications Magazine*, vol. 59, no. 6, pp. 112-118, June 2021,
DOI: 10.1109/MCOM.011.2100044 .

[2] H. P. Yuen, "KCQ: A new approach to quantum cryptography I. General principles and key generation," https://arXiv:quant-ph/0311061v6, 2003.

[3] M. Yoshida, T. Kan, K. Kasai, T. Hirooka and M. Nakazawa, "10 Tbit/s QAM Quantum Noise Stream Cipher Coherent Transmission Over 160 Km," *Journal of Lightwave Technology*, vol. 39, no. 4, pp. 1056-1063, 15 Feb.15, 2021,
DOI: 10.1109/JLT.2020.3016693 .

[4] K. Tanizawa and F. Futami, "Ultra-long-haul digital coherent PSK Y-00 quantum stream cipher transmission system," *Optics. Express*, vol. 29, no. 7, pp. 10451-10464, 2021.
DOI: 10.1364/OE.418302

[5] H. Jiao, T. Pu, J. Zheng, H. Zhou, L. Lu, P. Xiang, J. Zhao, and W. Wang, "Semi-quantum noise randomized data encryption based on an amplified spontaneous emission light source," *Optics Express*, vol. 26, no. 9, pp. 11587-11598, 2018,
DOI: 10.1364/OE.26.011587

[6] J. Cho and P. J. Winzer, "Probabilistic Constellation Shaping for Optical Fiber Communications," *Journal of Lightwave Technology*, vol. 37, no. 6, pp. 1590-1607, 15 March15, 2019,
DOI: 10.1109/JLT.2019.2898855 .

[7] J. Sun, L. Jiang, A. Yi, J. Feng, X. Deng, W. Pan, Bi. luo, and L. Yan, "Experimental demonstration of 201.6-Gbit/s coherent probabilistic shaping QAM transmission with quantum noise stream cipher over a 1200-km standard single mode fiber," *Optics Express*, vol. 31, no. 7, pp. 11344-11353, 2023,
DOI: 10.1364/OE.484431

[8] P. Schulte and G. Böcherer, "Constant Composition Distribution Matching," in *IEEE Transactions on Information Theory*, vol. 62, no. 1, pp. 430-434, Jan. 2016,
DOI: 10.1109/TIT.2015.2499181